# Prediction of electronic couplings for molecular charge transfer using optimally-tuned range-separated hybrid functionals

Debashree Manna,[1] Jochen Blumberger,[2,3] Jan M. L. Martin,[4] and Leeor Kronik[1]

1. Department of Materials and Interfaces, Weizmann Institute of Science, Reḥovoth 76100, Israel

2. Thomas Young Centre and Department of Physics and Astronomy, University College London, Gower Street, London WC1E 6BT, United Kingdom

3. Institute for Advanced Study, Technische Universitat Munchen, Lichtenbergstrasse 2 a, D-85748 Garching, Germany

4. Department of Organic Chemistry, Weizmann Institute of Science, Reḥovoth 76100, Israel

**Abstract**

Electronic coupling matrix elements are important to the theoretical description of electron transfer processes. However, they are notoriously difficult to obtain accurately from time-dependent density functional theory (TDDFT). Here, we use the HAB11 benchmark dataset of coupling matrix elements to assess whether TDDFT using optimally-tuned range-separated hybrid functionals, already known to be successful for the description of charge transfer excitation energies, also allows for an improved accuracy in the prediction of coupling matrix elements. We find that this approach outperforms all previous TDDFT calculations, based on semi-local, hybrid, or non-tuned range-separated hybrid functionals, with a remaining average deviation as low as ~12%. We discuss potential sources for the remaining error.

## Introduction

The concept of diabatic electronic states has proven to be highly successful, among other things, in the description of electron transfer reactions.[1–3] However, diabatic states are not eigenfunctions of the electronic Hamiltonian, but rather require the calculation of electronic off-diagonal elements (also known as electronic coupling matrix elements), denoted as $H_{ab}$:

$$H_{ab} = \langle \psi_a | \mathcal{H} | \psi_b \rangle, \qquad (1)$$

where $\mathcal{H}$ is the electronic Hamiltonian and $\psi_a, \psi_b$ are the diabatic states $a$ and $b$. In electron transfer, where only two (initial and final) diabatic states are considered, $H_{ab}$ takes a prominent role. This is because the electron transfer rate is proportional to $|H_{ab}|^2$ for small $H_{ab}$ values (non-adiabatic limit) and the electron transfer activation barrier is lowered by $H_{ab}$ for large $H_{ab}$ values (adiabatic limit).[1–3]

It is desirable to compute $H_{ab}$ accurately using computationally inexpensive methods, so that electron transfer properties of systems large enough to be of practical significance can be predicted from first principles. For a simple two-state donor-acceptor system, the adiabatic ground ($E_1$) and first excited state ($E_2$) potential energies are related to the diabatic potential energies $E_a$, $E_b$ by,

$$E_{2,1} = \frac{1}{2} \left( E_a + E_b \pm \sqrt{(E_a - E_b)^2 + 4|H_{ab}|^2} \right) \qquad (2)$$

Specifically, for symmetric systems, where $E_a = E_b$, the first adiabatic excitation energy, $\Delta E_{12} = E_2 - E_1$, is simply twice the desired matrix element $H_{ab}$, i.e.,

$$2|H_{ab}| = \Delta E_{12}. \qquad (3)$$

It is therefore tempting to use time-dependent density functional theory (TDDFT),[4,5] which has been used extensively to calculate excitation energies in molecules (see, e.g., Refs. 4–12 for selected overviews), in order to compute electronic coupling elements.

While TDDFT is a formally exact theory for molecular excited states, it is always approximate in practice. Practical success depends entirely on the accuracy of the approximate exchange-correlation functional employed in the calculations. Unfortunately, for charge-transfer excitations, which are essential to predicting electron transfer states, TDDFT using conventional approximations is well-known to fail, producing very large quantitative errors and even qualitative ones.[13–19]

Stein et al. have shown that charge transfer excitation energies can be reliably predicted from TDDFT, using the concept of the optimally-tuned range-separated hybrid (OT-RSH) functional.[20,21] In the RSH approach,[22–24] the repulsive Coulomb potential is range-split, allowing for the separate treatment of each interaction range. For the type of RSH functionals of interest here, the functional reduces to the Hartree–Fock approximation in the long range and to a semi-local or a conventional hybrid functional in the short-range. This allows for an asymptotically correct potential - which is essential to the correct description of charge transfer - while retaining a careful balance between short-range exchange and correlation - which is essential to the proper description of chemical bonds. In the optimal tuning approach,[25,26] one chooses the range-separation parameter (the value of which controls the transition between short- and long-range interactions) non-empirically, by enforcing (possibly for multiple charge states) the ionization potential theorem.[27–33]

In recent years, the OT-RSH approach has been used successfully to predict a wide range of charge-transfer phenomena (see, e.g., Refs. 18–21,34–37). However, to the best of our knowledge it has rarely been used for the calculation of electronic coupling[38] and its accuracy has not been systematically assessed. An excellent opportunity for such an assessment is afforded by the recently suggested HAB11 and HAB7 databases of Kubas et al.[39,40] The former database, on which we focus in the present study, consists of 11 π-conjugated organic homo-dimer cations, possessing different numbers of multiple bonds, varying types of aromaticity, and different heteroatoms (N, O, S). For each dimer cation, high-level *ab initio* benchmark data are provided for four different distances between the monomers forming the dimers.

In this article, we provide a comprehensive TDDFT investigation of the electronic coupling reported in the HAB11 dataset. We show that indeed TDDFT using conventional semi-local and hybrid functionals fails for these systems, that RSH functionals offer a distinct improvement, and that the OT-RSH approach results are the most accurate. Finally, we discuss remaining discrepancies and their possible origin.

**Computational Approach**

To explore how TDDFT performs for the calculation of the electronic coupling elements in the HAB11 database, we first use the well-known generalized-gradient-approximation (GGA) of Perdew, Burke, and Ernzerhof (PBE),[41] as well as the global hybrid functional based on it, PBE0,[42] with the usual exact-exchange fraction of 25% and also with an increased fraction of 50%. We then employ three standard (non-optimally-tuned) RSH functionals: CAM-B3LYP,[43] ωB97X[44], and LRC-ωPBEh.[45] Finally, we optimally-tune and utilize the LRC-ωPBEh functional, as explained below.

In RSH functionals, the coulomb potential is split into two terms using the identity[43]

$$\frac{1}{r} = \frac{\alpha + \beta \, \mathrm{erf}(\gamma r)}{r} + \frac{1 - (\alpha + \beta \, \mathrm{erf}(\gamma r))}{r} \quad , \tag{4}$$

where $\alpha$, $\beta$ and $\gamma$ are adjustable parameters, and $r$ is the inter-electron coordinate. The first term is treated using Harree-Fock exchange and the second term is treated using GGA exchange. GGA correlation is used throughout. The three RSH functionals we use differ on the choice of the three adjustable parameters, as well as on the choice of the GGA expressions used for the short-range exchange and the correlation.

For optimal tuning of the range-separation parameter $\gamma$, we rely on the ionization potential theorem. This theorem, obeyed by the exact functional, identifies the energy of the highest occupied orbital with (minus) the ionization potential obtained from total energy differences of the original system and the ionized one. Here, we employ this principle twice – for both the $N$ electron and the $N+1$ electron system, so as to minimize the deviation of the energy of the lowest unoccupied molecular orbital (LUMO) from the electron affinity of the system.[20,26,46–49] Thus, $\gamma$ is obtained by minimizing the target function $J^2(\gamma)$, given by

$$J^2(\gamma) = (\varepsilon^\gamma_{HOMO(N)} + I^\gamma(N))^2 + (\varepsilon^\gamma_{HOMO(N+1)} + I^\gamma(N+1))^2 \qquad (5)$$

where $\varepsilon^\gamma_{HOMO}$ is the energy of the highest occupied molecular orbital (HOMO) for a specific $\gamma$ and $I^\gamma$ is the ionization potential of the system, obtained from ground state energy differences between the original system and one where an electron has been removed, for the same $\gamma$ value. ($N$) and ($N+1$) denote the number of electrons in the system. We use the LRC-ωPBEh functional as the basis for the optimal tuning for several reasons. First, it uses $\alpha+\beta=1$, which guarantees use of 100% Fock exchange in the long-range, which is essential for the description of charge transfer.[20] Second, it uses $\alpha=0.2$, which means that 20% of Fock-exchange are used in the short range. We have previously found this to be a highly suitable value,[45,50–52] being in the range typically used by global hybrid functionals and therefore affording a good balance of exchange and correlation in the short-range. We note that the CAM-B3LYP functional does not use 100% Fock exchange at long range. The ωB97X functional does, but we preferred to tune the LRC-ωPBEh one as it contains fewer semi-empirical parameters. It has been previously reported that constrained DFT (CDFT) using a larger 50% fraction of Fock exchange yielded best agreement with reference values,[39] an issue elaborated below. Therefore, for comparison purpose we have also considered a modified PBE0 functional with 50% of Fock-exchange, as well as a modified optimally-tuned LRC-ωPBEh functional, with $\alpha=0.5$.

All molecular coordinates were those used in the construction of the HAB11 dataset (see supporting information of Ref. [39]). All TDDFT calculations were carried out within the linear response approach using QChem 4.3[53] and Gaussian 09[54] with the cc-pVTZ[55] basis set. We emphasize that the results reported below do *not* employ the Tamm-Dancoff approximation (TDA) to TDDFT.[56] While the TDA is often an excellent approximation to full-matrix linear-response TDDFT, in this case it has been found to introduce substantial errors, as shown in Table S1 of the supplementary material for the case of the PBE0 functional.

**Results and Discussion**

A thorough comparison of the TDDFT results for the electronic coupling elements, as obtained with the various exchange-correlation functionals discussed above, is reported in Table I and shown graphically (on a logarithmic scale) in Figure 1. Data are presented for ten out of the eleven systems in the dataset - the benzene dimer cation has been omitted owing to severe convergence issues of the TDDFT calculations. In addition to the $|H_{ab}|$ values, Table I and figure 1 also report the exponential decay constant of the coupling, related to $|H_{ab}|$ via

$$|H_{ab}| = A\ exp(-\beta d/2), \tag{6}$$

where $d$ is the inter-monomer distance. $\beta$ has been extracted by fitting the dependence of $|H_{ab}|$ on $d$ to an exponent, and is reported as n/a for functional and system combinations where the dependence of $|H_{ab}|$ on $d$ deviated significantly from being exponential. Finally, for convenience Table I reports the reference values and the TDDFT *deviations from it,* for both $|H_{ab}|$ and $\beta$. The Table also provides error statistics, via the mean unsigned error (MUE), mean relative signed error (MRSE), mean relative unsigned error (MRUE), and maximum unsigned error (MAX). could not see MAX in Table I.

An initial observation is that, as expected, TDDFT with GGA and conventional hybrid functionals performs very poorly. With PBE, the mean relative unsigned error (MRUE), in %, with respect to the reference values is ~208%, i.e., the results are in gross quantitative error. With PBE0, a global hybrid functional, severe quantitative deviations are somewhat improved, but the MRUE in % is still quite high, ~168%. Increasing the PBE0 percentage of Hartree-Fock exchange from 25% to 50%, a value typically too large for thermochemistry, reduces the deviation, but the MRUE is still a far from satisfying ~86%.

Table I and Figure 1 additionally show that errors are further reduced by using any of the three RSH functionals tested, but not to the same degree. CAM-B3LYP and LRC-ωPBEh offer a more modest improvement compared to the PBE and PBE0 functionals, reducing the deviation to ~111% and ~88% respectively, with LRC-ωPBEh being moderately but consistently more accurate. The best improvement is obtained using ωB97X, with the errors reduced to ~34%. Still, for some of the systems deviations are noticeably larger (e.g., acetylene, cyclopentadiene) and some calculations were unsuccessful due to severe convergence problems (e.g., furan, pyrrole).

The relatively modest improvement obtained from CAM-B3LYP can be easily rationalized by considering that this functional, fit against thermochemistry data, only contains 65% Fock exchange in the long-range. While this is more than PBE0, it still falls short of the 100% needed to obtain the correct asymptotic potential. It is also easy to rationalize the relative success of the ωB97X functional (fit against thermochemistry, kinetics, and non-covalent interaction data), which does possess 100% Fock exchange asymptotically. It is less apparent, at first glance, why LRC-ωPBEh, which also possesses full asymptotic Fock exchange, performs more like CAM-B3LYP than like ωB97X, a point we revisit below.

Next, we study the effect of optimal tuning, using Eq. (5), on the LRC-ωPBEh results. Two different varieties of optimal tuning were used. In one approach, denoted as OT-RSH-d, optimal tuning is performed directly on the dimer system, which means that it is performed separately for each chemical system at each monomer distance. In another approach, denoted as OT-RSH-m, optimal tuning is performed on the neutral monomer and the range-separation parameter thus determined is used for the dimer at all monomer separation. This means that the procedure is performed only once per each chemical system. Optimal $\gamma$ values obtained from the two methods for all systems studied, using the default short-range exchange fraction of $\alpha=0.2$, are given in Table S2 of the Supporting Information. Clearly, in either its "d" or "m" variant, optimal tuning greatly improves the LRC-ωPBEh results. In fact, the OT-RSH-d results are statistically comparable to those of ωB97X and OT-RSH-m results are even slightly better, with MRUEs of 11.3% and 11.7% for $H_{ab}$ and $\beta$ values, respectively.

Clearly, optimal tuning has a decisive effect on the quality of the results. In other words, incorporation of asymptotic exact exchange is, in and of itself, not a sufficient condition for obtaining quality results. Rather, the coupling energies are also quite sensitive to the precise value of the range-separation parameter $\gamma$. For the ωB97X functional, the default $\gamma$ value, 0.3 bohr$^{-1}$, happens to be close to the optimally-tuned one, explaining its success (with a similar default value used in other semi-empirical range-separated hybrid functional[43,57]). For the LRC-ωPBEh functional, the default value of $\gamma$ is a smaller 0.2 bohr$^{-1}$, which is too small for these systems. Therefore, it benefits substantially from the tuning procedure. Generally, it has been demonstrated repeatedly that the optimal $\gamma$ value can be a strong function of the system size and chemical composition.[47,58–63] Therefore, optimal-tuning is highly recommended for a general system outside the specific HAB11 dataset.

We note that for some data points – notably the acetylene dimer cation at all inter-monomer distances and the thiophene dimer cation at a 3.5 Å and 4.0 Å inter-monomer distances – the lowest energy excitation is not the charge transfer one, i.e., not the excitation to the singly occupied orbital.[39] Therefore, it is essential to consider the right excitation when comparing to the HAB11 reference data. We further note that in these cases the tuning process was often of lower quality. To understand this, Fig. 2 shows a plot of the tuning target, $J^2(\gamma)$ (see Eq. (5)) as a function of the range-separation parameter γ, for two cation dimers, the ethylene dimer (a) and the acetylene dimer (b), both at an inter-monomer separation of 3.5 Å. For ethylene, a clear and relatively deep minimum at a value close to zero is observed. For acetylene, however, that is not the case – the minimum is shallow and differs from zero substantially. Failure of the tuning procedure is not common but is not unprecedented.[64] It simply means that in difficult cases, e.g., strong heterogeneity or more complicated correlation, the functional form is not flexible enough to find a range-separation parameter for which the ionization potential theorem is fully obeyed. This serves, then, as a "built-in indicator" that results would not be as reliable. Nevertheless, in our case the performance is still quite satisfactory. For the overwhelming majority of cases studied here, this problem was not encountered.

At this point, it is of interest to compare the electronic coupling and decay constant values calculated here with those obtained previously[39,65] using methods based on ground-state DFT, namely, CDFT,[66] fragment-orbital DFT (FODFT),[67] self-consistent charge density functional tight-binding (FODFTB),[68] and projector operator-based diabatization (POD)[65] It was found that both CDFT and POD with 50% Fock exchange yielded the best agreement to the reference values, with a 5.3% and 9.3% MRUE for the $H_{ab}$, a distinct improvement over the 13.8% and 17.1% MRUE obtained with CDFT and POD using 25% Fock exchange. We therefore examined the effect of increasing the short-range fraction of Fock exchange in OT-RSH-d to 50%, while re-optimizing γ. We found that, compared to OTRSH-d with 20% Fock exchange, this somewhat further decrease the MRUE for $H_{ab}$ to 15.3%, with an MRUE of 14.8% for $β$ values. This error is still larger than that found by CDFT, but several arguments stand in its favour. First, this OT-RSH result outperforms that obtained with GGA, hybrid, or other range-separated hybrid functionals. Second, the quality of the TDDFT-based calculations generally does not depend on the inter-monomer separation, whereas the results of CDFT calculations for small inter-monomer distances can depend on the choice of the weight functions used for constraining the charge. Third, it is important to note that the errors made by the OTRSH-d and OTRSH-m are of the order of few tens of meV, which is in fact better than the accuracy obtained with OT-RSH approaches for electronic or optical excitations (typically 0.1 eV to 0.2 eV).

Finally, because we are dealing with small, symmetric systems, one could conjecture that much of the remaining discrepancy between TDDFT and reference values arises from static correlation. However, two arguments stand against this conjecture. First, as noted above overall accuracy somewhat improves with an increasing fraction of short-range exchange, whereas the opposite is true in the presence of strong static correlation.[69–71] Second, Fogueri et al.[72] suggested a DFT-based diagnostic for nondynamical correlation, given by $A_\lambda = (1-TAE[X_\lambda C]/TAE[XC])/\lambda$, where TAE is the molecular total atomization energy, XC represents a pure-DFT exchange-correlation functional, and $X_\lambda C$ represents the corresponding hybrid with $100\lambda$ % Fock exchange. $A_\lambda$ values around or above 1 indicate presence of severe static correlation, $A_\lambda$ values near 0.15 indicate mild static correlation, and $A_\lambda$ values below 0.10 point indicate mostly dynamic correlation. For all of our systems, $A_\lambda$ values were found to be between 0.070 and 0.134, supporting the absence of strong static correlation.

**Conclusion**

In conclusion, we have used the recently developed HAB11 benchmark dataset to assess the performance of TDDFT in computing electronic coupling matrix elements in small cationic dimers. We compared semi-local functionals, global hybrid functionals, conventional range-separated hybrid functionals, and optimally-tuned range-separated hybrid functionals. We found that the latter decisively provide the best overall agreement with benchmark data, to within ~12%. Future challenges for further reduction of the deviation between TDDFT and benchmark data include an *a priori* identification of the optimal fraction of short-range exact exchange and further improvements in correlation approximations. Moreover, since anions are often more challenging than cations in (TD)DFT, it will be of interest to test the performance of OTRSH on the HAB7 database of coupling matrix elements for dimer-anions.

**Acknowledgement**
DM acknowledges partial financial support from the Feinberg Graduate School of the Weizmann Institute of Science. JB acknowledges the European Union for cofunding of a TUM-IAS Hans-Fischer Fellowship. LK acknowledges the historical generosity of the Perlman Family. This research was partially supported by the Israel Science Foundation (grant 1358/15 to JMLM).

**Supporting Information:** Deviations from the reference values for the TDDFT excitation energies with the PBE0 functional, with and without the Tamm-Dancoff approximation are provided in Table S1. Optimal γ values for the OTRSH-m and OT-RSH-d methods (using α=0.2) are provided in table S2. This material is available free of charge via the Internet at http://pubs.acs.org.

**TABLE I.** Reference values for electronic coupling matrix elements $H_{ab}$ (in meV), and deviations from the reference values obtained with different functionals, for 10 out of the 11 dimer cations in the HAB11 data set, at various distances between monomers. Reference values and deviations are also given for the decay constant $\beta$ (in 1/Å). Also provided are the mean unsigned error [MUE = $(\Sigma_n |y_{cal}-y_{ref}|)/n$], the mean relative signed error [MRSE = $(\Sigma_n ((y_{cal}-y_{ref})/y_{ref}))/n$], the mean relative unsigned error [MRUE = $(\Sigma_n (|y_{cal}-y_{ref}|/y_{ref}))/n$] and the maximum unsigned error (MAX = max $|y_{calc}-y_{ref}|$) for $H_{ab}$ and $\beta$.

| Dimer cations | | Distance between monomers (Å) | REF[a] | OTRSH-d with α=0.2 | OTRSH-d with α=0.5 | OTRSH-m with α=0.2 | PBE | PBE0 with 25% HF | PBE0 with 50% HF | CAM-B3LYP | LRC-ωPBEh | wB97X |
|---|---|---|---|---|---|---|---|---|---|---|---|---|
| Ethylene | $H_{ab}$ | 3.50 | 519.20 | 60.60 | 7.75 | 9.65 | 418.25 | 301.20 | 119.10 | 204.45 | 187.60 | 42.85 |
| | | 4.00 | 270.80 | 51.35 | 13.55 | 10.15 | 387.35 | 298.45 | 142.55 | 213.95 | 188.75 | 52.45 |
| | | 4.50 | 137.60 | 42.20 | 19.25 | 13.90 | 328.15 | 264.10 | 146.30 | 196.60 | 171.25 | 56.60 |
| | | 5.00 | 68.50 | 30.65 | 20.05 | 14.30 | 254.95 | 211.05 | 129.15 | 162.65 | 138.95 | 48.90 |
| | β | | 2.70 | -0.45 | -0.32 | -0.23 | -1.39 | -1.27 | 1.14 | -1.18 | -1.07 | -0.62 |
| Acetylene | $H_{ab}$ | 3.50 | 460.70 | 119.15 | 61.90 | 12.30 | 463.05 | 346.60 | 163.15 | 257.55 | 247.15 | 108.20 |
| | | 4.00 | 231.80 | 88.35 | 64.25 | 12.95 | 408.70 | 323.60 | 179.85 | 251.55 | 234.55 | 116.40 |
| | | 4.50 | 114.80 | 71.20 | 57.00 | 13.75 | 329.70 | 270.90 | 166.10 | 217.20 | 198.55 | 104.60 |
| | | 5.00 | 56.60 | 50.30 | 39.80 | 14.10 | -- | 209.80 | 137.70 | 171.15 | 145.35 | 85.40 |
| | β | | 2.80 | -0.55 | -0.55 | -0.26 | -1.34 | -1.32 | -1.25 | -1.27 | -1.14 | -0.95 |
| Cyclopropene | $H_{ab}$ | 3.50 | 536.60 | 34.65 | -20.9 | 3.05 | 373.35 | 274.15 | 97.30 | 156.95 | 155.10 | 9.80 |
| | | 4.00 | 254.00 | 36.25 | -3.35 | 17.45 | 371.85 | 280.70 | 132.25 | 180.80 | 167.40 | 37.50 |
| | | 4.50 | 118.40 | 34.05 | 11 | 28.70 | 305.90 | 244.15 | 136.75 | 169.05 | 152.50 | 41.55 |
| | | 5.00 | 54.00 | 26.40 | 15.95 | 30.70 | 233.10 | 191.70 | 119.35 | 139.40 | 123.40 | 37.55 |
| | β | | 3.06 | -0.45 | -0.40 | -0.59 | -1.52 | -1.47 | -1.34 | -1.37 | -1.25 | -0.68 |
| Cyclobutadi | $H_{ab}$ | 3.50 | 462.70 | 39.00 | -8.25 | -12.20 | 313.90 | 229.40 | 89.05 | 114.30 | 104.70 | -24.75 |

| | | | | | | | | | | | |
|---|---|---|---|---|---|---|---|---|---|---|---|
| ene | | 4.00 | 239.10 | 36.40 | 5.25 | 1.50 | 314.35 | 235.60 | 114.15 | 137.00 | 118.60 | -7.00 |
| | | 4.50 | 121.70 | 36.00 | 16.45 | 13.10 | 262.75 | 213.80 | 120.85 | 135.65 | 115.25 | 0.85 |
| | | 5.00 | 62.20 | 31.60 | 18.35 | 16.50 | 206.50 | 173.25 | 107.40 | 116.40 | 96.85 | 2.25 |
| | β | | 2.68 | -0.44 | -0.38 | -0.35 | -1.26 | -1.25 | -1.11 | -1.12 | -0.99 | -0.12 |
| | | 3.50 | 465.80 | 14.60 | -55.1 | -41.45 | 274.15 | 183.40 | 28.60 | 62.60 | 46.15 | -114.85 |
| Cyclopenta diene | H$_{ab}$ | 4.00 | 234.40 | 31.75 | -30.15 | -19.75 | 286.70 | 204.40 | 67.00 | 98.00 | 68.95 | -113.15 |
| | | 4.50 | 114.30 | 43.65 | -5.7 | -1.35 | 260.75 | 193.90 | 90.35 | 108.70 | 80.05 | --[b] |
| | | 5.00 | 53.40 | 39.20 | 9.85 | 7.75 | 214.35 | 161.15 | 89.85 | 100.60 | 74.45 | --[b] |
| | β | | 2.89 | -0.70 | -0.39 | -0.30 | -1.54 | -1.42 | -1.25 | -1.25 | -1.05 | -- |
| | | 3.50 | 440.30 | 31.80 | -35.45 | -28.85 | 297.10 | 208.75 | 61.05 | 96.45 | 78.55 | -62.90 |
| Furane | H$_{ab}$ | 4.00 | 214.90 | 38.95 | -12.3 | -8.60 | 297.40 | 222.80 | 88.85 | 126.10 | 101.15 | -42.85 |
| | | 4.50 | 101.80 | 47.75 | 9 | 7.10 | 257.50 | 205.00 | 104.00 | 127.65 | 103.55 | -40.55 |
| | | 5.00 | 46.00 | 40.45 | 18.30 | 14.15 | 202.40 | 167.70 | 97.55 | 112.00 | 90.15 | -- |
| | β | | 3.01 | -0.76 | -0.56 | -0.45 | -1.56 | -1.54 | -1.35 | -1.38 | -1.23 | 0.63 |
| | | 3.50 | 456.30 | 20.60 | -47.7 | -31.90 | 281.70 | 193.35 | 32.20 | 64.80 | 59.00 | -88.00 |
| Pyrrole | H$_{ab}$ | 4.00 | 228.60 | 37.55 | -24.85 | -13.15 | 294.35 | 210.60 | 69.20 | 109.60 | 81.70 | -73.25 |
| | | 4.50 | 111.30 | 45.35 | -0.9 | 3.10 | 255.10 | 198.15 | 90.30 | 116.75 | 89.10 | --[b] |
| | | 5.00 | 52.20 | 38.15 | 13.10 | 11.25 | 204.20 | 165.45 | 89.55 | 105.90 | 80.75 | --[b] |
| | β | | 2.89 | -0.68 | -0.44 | -0.36 | -1.48 | -1.44 | -1.25 | -1.30 | -1.09 | -- |
| | | 3.50 | 449.00 | 18.50 | -122.25 | -40.15 | 286.55 | 232.90 | 11.35 | --[b] | 39.70 | -117.80 |
| Thiophene | H$_{ab}$ | 4.00 | 218.90 | 38.95 | -35.05 | -22.70 | 285.05 | 203.70 | 51.00 | --[b] | 64.50 | -139.55 |
| | | 4.50 | 106.50 | 41.60 | -13.90 | -11.20 | 248.25 | 189.35 | 72.65 | --[b] | 71.10 | -- |
| | | 5.00 | 54.40 | 32.35 | -1.40 | -1.40 | 192.10 | 152.50 | 70.90 | --[b] | 61.10 | -- |
|  |  | β |  | 2.82 | -0.58 | -0.37 | -0.08 | -1.37 | -1.25 | -1.10 |  | -0.90 | -- |
|  |  |  | 3.50 | 411.60 | 52.80 | -11.1 | 0.05 | 327.00 | 237.60 | 82.80 | 126.10 | 110.40 | -28.75 |
| **Imidazole** | H$_{ab}$ | 4.00 | 202.80 | 59.70 | 0.15 | 6.05 | 318.30 | 236.00 | 98.60 | 140.60 | 117.55 | -22.05 |
|  |  | 4.50 | 99.10 | 47.65 | 11.25 | 11.75 | 275.95 | 209.10 | 105.55 | 132.60 | 109.50 | -26.35 |
|  |  | 5.00 | 49.70 | 32.90 | 14.50 | 12.00 | 217.95 | 164.85 | 93.55 | 110.20 | 89.00 | -- |
|  | β |  | 2.82 | -0.52 | -0.38 | -0.29 | -1.47 | -1.35 | -1.18 | -1.21 | -1.06 | 0.50 |
|  |  | 3.50 | 375.00 | 31.40 | -35.1 | -19.25 | 241.10 | 170.85 | 35.10 | 60.15 | 40.65 | -93.45 |
| **Phenol** | H$_{ab}$ | 4.00 | 179.60 | 37.70 | -17.75 | -2.00 | 244.00 | 180.55 | 60.90 | 85.70 | 58.80 | -104.80 |
|  |  | 4.50 | 85.20 | 40.60 | 1.6 | 10.25 | 203.10 | 163.70 | 74.70 | 89.40 | 63.95 | -- |
|  |  | 5.00 | 41.30 | 32.40 | 11.3 | 14.15 | 157.00 | 131.70 | 70.40 | 78.50 | 55.75 | -- |
|  | β |  | 2.95 | -0.68 | -0.46 | -0.47 | -1.44 | -1.42 | -1.23 | -1.23 | -1.02 | -- |
|  |  | **MUE (meV)** | 42.11 | 23.02 | 14.09 | 284.46 | 216.40 | 95.93 | 135.85 | 109.54 | 61.50 |
| |H$_{ab}$| | **MRSE (%)** | 34.48 | 7.59 | 7.17 | 208.31 | 168.41 | 86.20 | 111.44 | 88.32 | 5.24 |
|  |  | **MRUE (%)** | 34.48 | 15.26 | 11.33 | 208.31 | 168.41 | 86.20 | 111.44 | 88.32 | 33.52 |
|  |  | **MAX (meV)** | 119.15 | 122.25 | 41.45 | 463.05 | 346.60 | 179.85 | 257.55 | 247.15 | 139.55 |
|  |  | **MUE (1/Å)** | 0.58 | 0.42 | 0.34 | 1.44 | 1.37 | 1.22 | 1.26 | 1.08 | 0.58 |
| β | **MRSE (%)** | -20.22 | -14.80 | -11.70 | -50.17 | -47.90 | -42.59 | -43.79 | -37.70 | -7.49 |
|  |  | **MRUE (%)** | 20.22 | 14.80 | 11.70 | 50.17 | 47.90 | 42.59 | 43.79 | 37.70 | 20.33 |
|  |  | **MAX(1/Å)** | 0.76 | 0.56 | 0.59 | 1.56 | 1.54 | 1.35 | 1.38 | 1.25 | 0.95 |

[a] Ethylene, acetylene, cyclopropene, cyclobutadiene, cyclopentadiene, furane, and pyrrole reference values are calculated at the MRCI+Q level of theory. Thiophene, imidazole, benzene, and phenol reference values are calculated at the NEVPT2 level of theory.

[b] Values are not reported due to convergence problem.

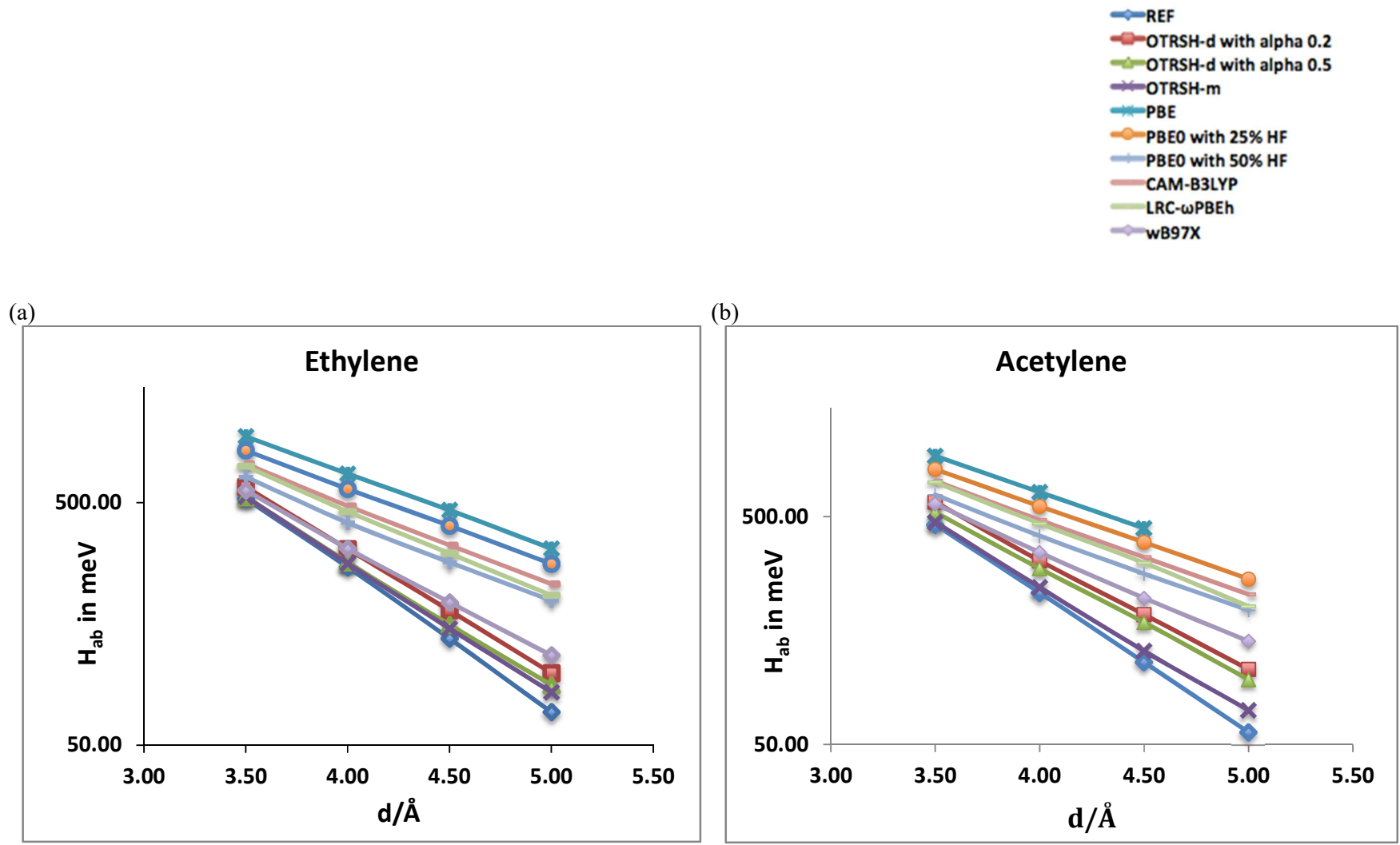

**Figure 1.** Logarithmic plot of |H$_{ab}$| (in meV) as a function of *d*, the inter-monomer distance, in Å, for all systems studied in this article.

(c)
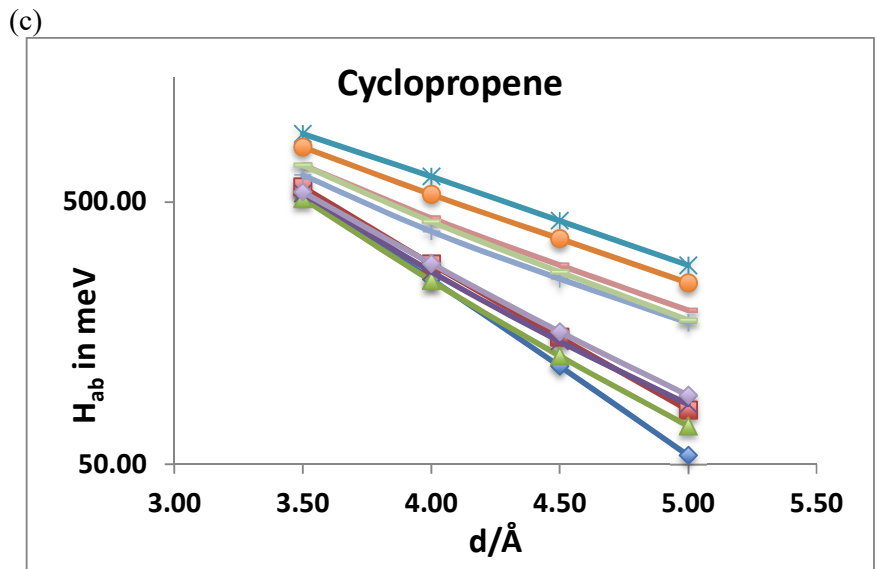

(d)
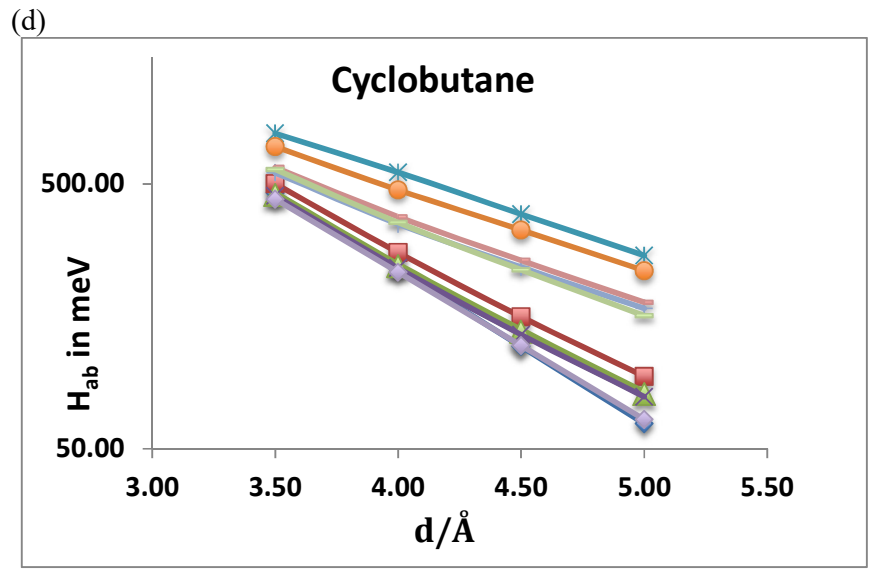

(e)
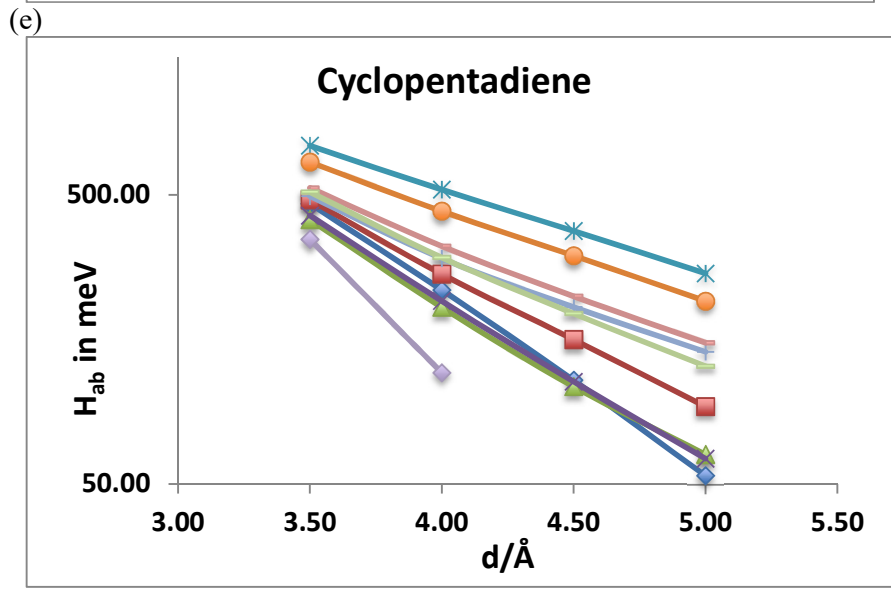

(f)
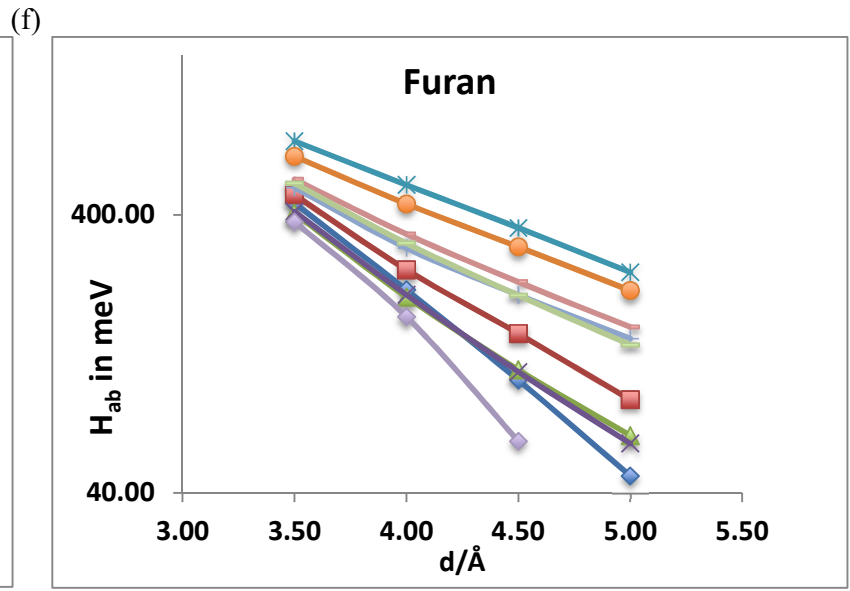

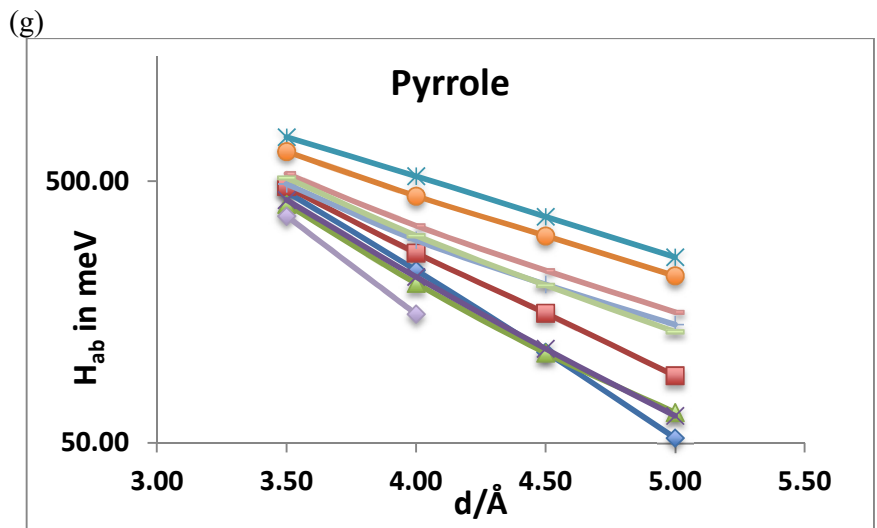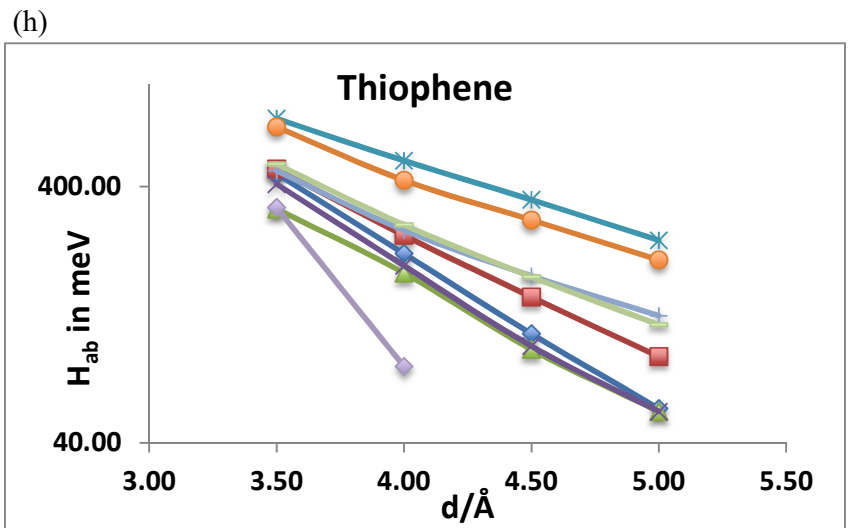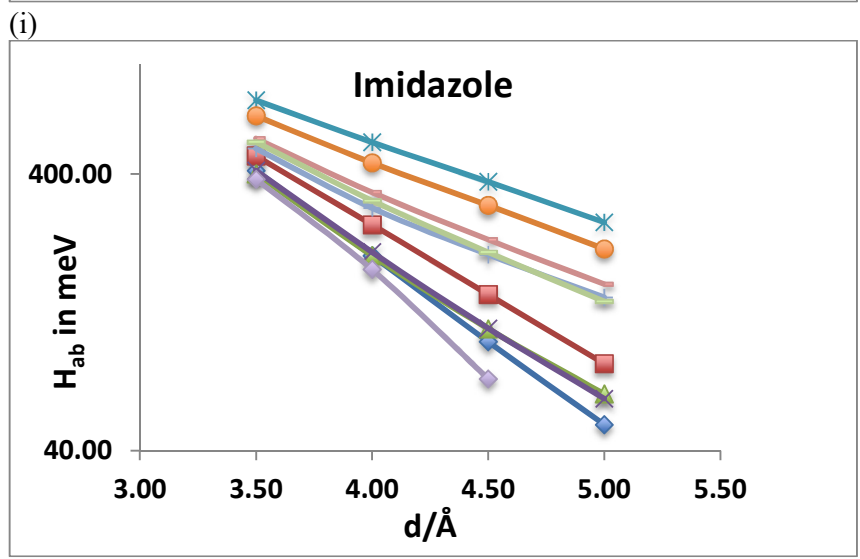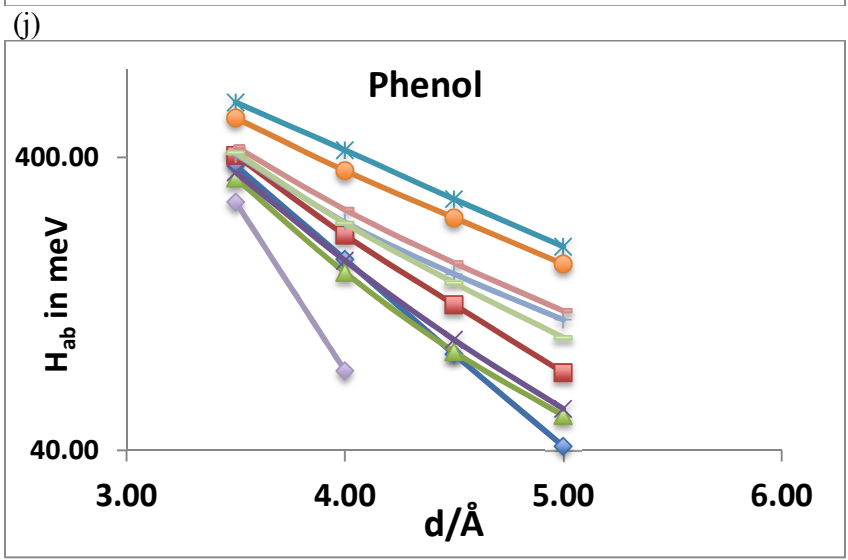

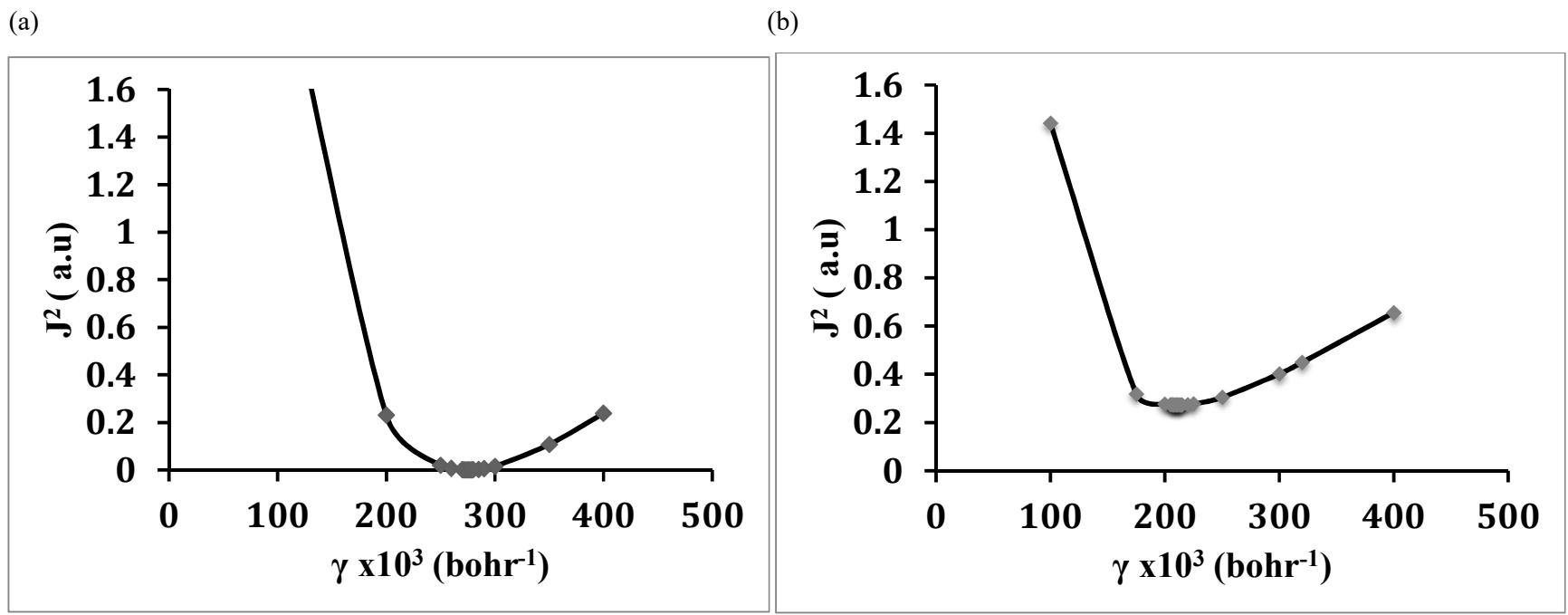

**Figure 2.** Plot of $J^2(\gamma)$ (see Eq. (5)) as a function of the range-separation parameter $\gamma$, for two cation dimers: (a) the ethylene dimer, (b) the acetylene dimer, both at an inter-monomer distance of 3.5 Å.